\documentclass[journal]{IEEEtran}
\usepackage[utf8x]{inputenc} 
\usepackage{contour}
\usepackage{verbatim}
\usepackage{blindtext}
\usepackage{float}
\usepackage{calc}
\def\BibTeX{{\rm B\kern-.05em{\sc i\kern-.025em b}\kern-.08em    T\kern-.1667em\lower.7ex\hbox{E}\kern-.125emX}}
\usepackage{graphicx}
\usepackage{epstopdf}
\begin{document}

\title{Deconstruction and reconstruction of image-degrading effects in the human abdomen using Fullwave: phase aberration, multiple reverberation, and trailing reverberation.}

\author{Danai Eleni Soulioti, Francisco Santibanez,
        and~Gianmarco Pinton
\thanks{D. E. Soulioti, F. Santibanez  and G. Pinton are with Joint Department of Biomedical Engineering of the University of North Carolina at Chapel hill and North Carolina State University.}
}

\maketitle

\begin{abstract}
Ultrasound image degradation in the human body is complex and occurs due to the distortion of the wave as it propagates to and from the target. Here, we establish a simulation based framework that deconstructs the sources of image degradation into a separable parameter space that includes phase aberration from speed variation,  multiple reverberations, and trailing reverberation. These separable parameters are then used to reconstruct images with known and independently modulable amounts of degradation using methods that depend on the additive or multiplicative nature of the degradation. Experimental measurements and Fullwave simulations in the human abdomen demonstrate this calibrated process in abdominal imaging by matching relevant imaging metrics such as phase aberration, reverberation strength, speckle brightness and coherence length. Applications of the reconstruction technique are illustrated for beamforming strategies (phase aberration correction, spatial coherence imaging), in a standard abdominal environment, as well as in impedance ranges much higher than those naturally occurring in the body. 
\end{abstract}

\IEEEpeerreviewmaketitle

\section{Introduction}
In transabdominal human ultrasound imaging, many of the factors that contribute to poor image quality can be attributed to the abdominal wall. These factors can be easily observed in overweight and obese individuals who have thicker layers of subcutaneous fat than a normal-weight person as well as more fascial or connective tissue layers that supports this additional fat~\cite{bmiUS}. These additional subcutaneous tissue layers are the main contributors to aberration and reverberation of ultrasound waves. We have published many studies demonstrating that both phase aberration and reverberation play a significant role in degrading image quality~\cite{pinton2010,pinton2011,pinton2014}. In previous work ~\cite{pinton2011}, we analyzed the role of phase aberration and reverberation as two major mechanisms in the degradation of ultrasound image quality. In this analysis, we found that the two mechanisms had a roughly equivalent contribution in degrading image quality, and that their combined effects were responsible for the vast majority of image-quality degradation. We also hypothesized that reverberation clutter from subcutaneous tissue layers limited the effectiveness of phase aberration correction techniques by corrupting the RF channel signals. 

Aberration occurs when the speed of sound in tissue deviates from the constant value used in ultrasound beamforming (typically 1540 m/s) and distorts the ultrasound wave. The wave front distortion decreases  image  resolution  and  introduces  noise  that  obscures  the  target of  interest. Although  a  wide  variety of techniques have been proposed to compensate for aberration, many of the near-field correction techniques~\cite{flax88,nock89,ng94} have shown limited improvements in image quality ~\cite{trahey90,rigby00,dahl06}, and distributed aberration correction techniques ~\cite{fink92,liu94,fink96} are not practical for real-time implementation because they require point reflectors, which are not commonly found in tissue. Acoustic refraction, a component of aberration, leads to geometric distortions if not accounted for. Jaeger $et$ $al$~\cite{jaeger}, showed that for moderate speed-of-sound variations, refraction can be mathematically described within the concept of straight ray propagation, and is mainly manifested in the receive (Rx) delays during beamforming, rather than Tx aberrations that appear to have an insignificant effect on the lateral position of echoes.

Multiple reverberations occur when an acoustic wave gets reflected multiple times between layers of tissue of different properties. Reverberation clutter is often is attributed to these reverberation among tissue layers, scattering from off-axis targets or side and grating lobes, and aberrations. It manifests in the blurring of the image and along with aberration are the main degrading mechanisms responsible for poor image quality~\cite{pinton2012}. Inter-layer reflections that are transmitted in the direction of pulse propagation instead of returning to the transducer surface, add a long, low-amplitude tail to the pulse causing it to appear lengthened, resulting in additional clutter further away from the transducer and compromising resolution~\cite{pinton2012}.

The generation of an ultrasound image of the soft tissue in the body relies on the physics of acoustic wave propagation:  diffraction,  reflection,  scattering,  frequency dependent attenuation and non linearity. A direct simulation of the propagation to a target,  reflection,  and then propagation back to the transducer is computationally costly due to two fundamental physical scales:  i) the propagation distance($∼100\lambda$) and  ii) sub-resolution scatterers ($<\lambda/10$). To represent both length scales in 3D,  simulation fields with large number of points in space ($∼10^9$) are required. Furthermore, numerical methods are challenged by the extremely high dynamic range because the backscattered wave may be 100 dB smaller than the transmitted pulse. Therefore any numerical errors from the large amplitude forward propagating wave must be 100 dB smaller than the backscattered wave. Given these challenges, it is unsurprising that simulations of ultrasound images make approximations that remove the need to model propagation directly. The most widely used simulation tool, Field II, for example, relies on the spatial impulse response and Hankel functions to replace propagation with a convolution approach~\cite{fieldII}. However,  by  not  simulating  wave  propagation  directly,  effects  such  as  multiple  scattering,  reverberation,  distributed aberration,  and non linearity,  which can be determining factors of ultrasound image quality,  cannot be modeled.  There are a number of simulation tools that model nonlinear propagation ~\cite{sim91,sim03,sim11} and they have been used successfully in modeling therapeutic acoustic fields.  However they cannot model multiple reflections and backpropagation, which are necessary for imaging.  Ultrasonic propagation through fine scale heterogeneities has  been  simulated  previously  with  a  finite  difference  time  domain  (FDTD)  solution  of  the  2D  and  3D  linear wave equation~\cite{sim97,sim02}, but they use low order stencils that generate forward propagation errors that are large compared to the backscattered wave. Other simulation tools also exist that can model nonlinear ultrasound propagation based on spectral domain or k-space methods~\cite{sim10}.  These numerical tools can model the fundamental wave physics of backpropagation, non linearity, and attenuation. However, since the Fourier transform of spatially localized structures has broad support in the Fourier domain, spectral methods do not accurately model the low amplitude reflections from small structures that are a key component of ultrasound images.

In this study, the previous validated Fullwave simulation tool is used for a 2D study of the effects of the human abdomen in ultrasound imaging at 3.7~MHz. Since its first publication~\cite{pinton2009}, Fullwave has been updated and into a higher-order version as shown in recent work~\cite{fullwave2}.

\begin{figure}[ht]
\setlength{\unitlength}{0.5\textwidth}
\begin{picture}(0.5,0.5)(0,0)
\put(0.0,0.0){\includegraphics[trim= 0 0 20 0,clip,height=0.48\unitlength]{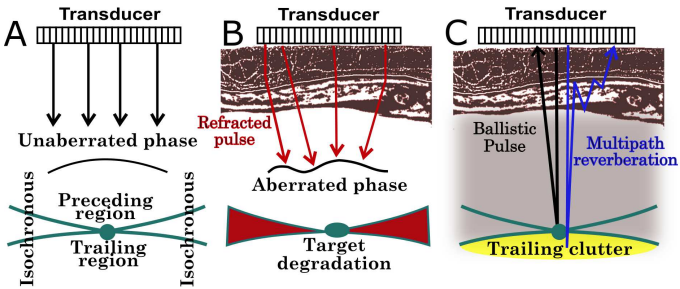}}
\end{picture}
 \caption{\label{fig:aberration_reverberation} A, in a homogeneous medium, a typical PSF is characterized by three different regions, preceding, trailing and the isochronous volume. In B, aberration degrades ultrasound focusing capability and compromised resolution. In C, multipath reverberation overlays acoustical noise to the echoes produced from the ballistic pulse, thereby reducing detectability. The lengthening of the pulse results in the appearance of trailing clutter. Both reverberation and aberration depend on body wall anatomy.}
\end{figure}

\section{Methods}
We utilized a C5-2v abdominal imaging transducer connected to a Vantage 256 ultrasound research scanner (Verasonics, Inc., Kirkland, WA) to capture RF channel signals from an abdominal phantom. The phantom comprised a thick slice or porcine abdomen, followed by a homogeneous slice of liver. These measurements are used to calibrate and validate the Fullwave simulations in terms of reverberation strength,  phase aberration,  speckle brightness and LOC. Reverberation curves are acquired by laterally averaging the beamformed RF data and plotting them as a function of depth. 
In previous work, we have used and characterized a 2-D heterogeneous tissue model of a human abdominal layer \cite{pinton2011}. The discrete structures in the model are assigned one of three tissue types: fat, muscle, or connective tissue and each type is given tabulated values for speed of sound, density, non-linearity, and attenuation \cite{goss1978} and are summarized in Table I. The abdominal section was also modified spatially to simulate compression in a clinical setting, where the ultrasonographer would apply pressure to the probe to adjust the image quality. 

\begin{figure}[ht]
\setlength{\unitlength}{0.7\textwidth}
\begin{picture}(0.5,0.8)(0,0)
\put(0.0,0.0){\includegraphics[trim= 0 0 0 0,clip,height=0.8\unitlength]{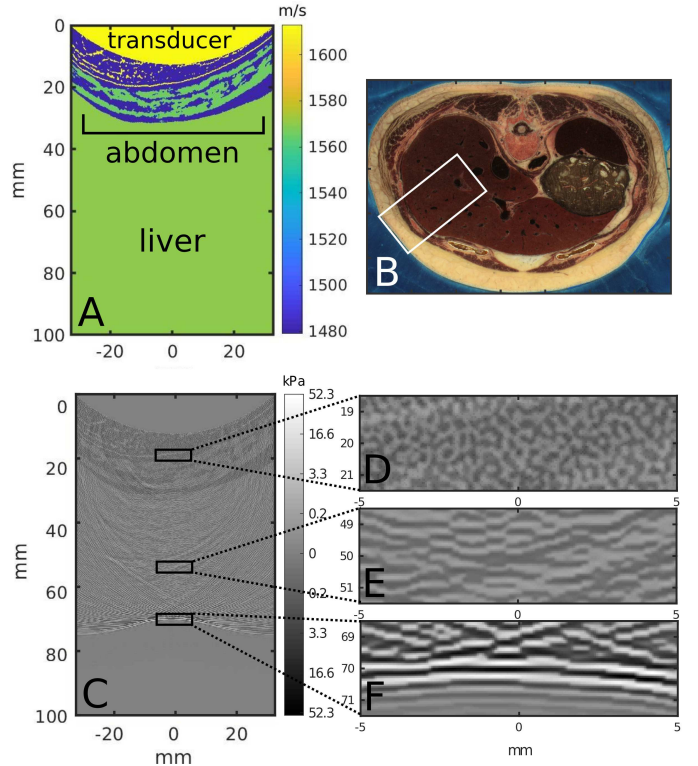}}
\end{picture}
\caption{(A) Example of an input sound speed map for Fullwave simulations, as segmented and processed from the selected portion of (B), the visible female abdominal dataset. In (C), the result of a wave propagating through map (A) is shown. Highlighted are different characteristic regions of propagation: (D) is the highly incoherent portion of the wave as it travels through the abdominal wall. As the wave enters the preceding region in (E), coherence increases as it approaches its natural focal zone. In (F) the wave has reached its maximum coherence value at the focus. }
\label{fig:cmap}
\end{figure}

All simulations were performed on a Linux Fedora 25 (v.4.10.13-200.fc25.x86 64) system running Intel Xeon\textsuperscript{\textregistered} E5-2630 v4  Processors at 2.20~GHz. The simulation code was written in C and post-processing was performed on MATLAB. Each simulation for each individual emission case has an approximate duration of 30~min.
The imaging medium is described with a spatial resolution of 34.7 um. To simulate a scattering ultrasonic medium, the medium is populated with point scatterers with a density of twelve scatterers per resolution cell. The point scatterers have a 34.7 um diameter and random spatial distribution. Their amplitude is defined by its difference in speed of sound from the surrounding medium, with the mean variation in the speed of sound corresponding to a 6\% variation of the accepted average tissue velocity of 1540 m/s.

\begin{figure}[h!]
\setlength{\unitlength}{0.8\textwidth}
\centering
\begin{picture}(0.5,0.8)(0,0)
\put(0.0,0.0){\includegraphics[trim= 0 0 0 0,clip,height=0.8\unitlength]{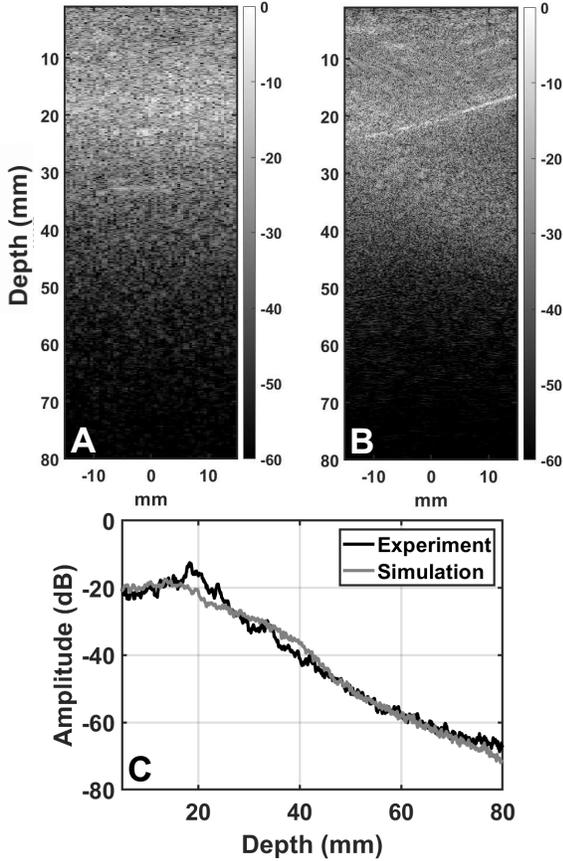}}
\end{picture}
\caption{A, Experimental B-mode using the C5-2v probe. B, simulated B-mode. C, comparison of experimental and simulation reverberation curves for the first 80~mm of depth.}
\label{fig:bmodes}
\end{figure}

To ensure proper calibration between the experimental case and our simulation tool, the RMS value of the aberration for the experiment was calculated at 145~ns and for the simulation at 138~ns, using correlation on two different phase profiles, in both experimental and simulation data: A target in a homogeneous medium (water) and a target under the abdomen at the exact same position. Experimentally this was achieved by multiple acquisitions prior and after inserting the pork belly and liver components. The same median frequency of 3.7~MHz and a plane wave emission is used in both experiment and simulations. Using Fullwave, we are able to model virtually any possible transducer shape and configuration, in both 2D and 3D. Since the C5-2v transducer modeled is curved, its geometry is represented as shown is Fig.~\ref{fig:cmap}A by the yellow arc portion.

\begin{figure}[h!]
\setlength{\unitlength}{0.8\textwidth}
\begin{picture}(0.5,0.45)(0,0)
\put(0.01,0.0){\includegraphics[trim= 0 0 0 0,clip,height=0.45\unitlength]{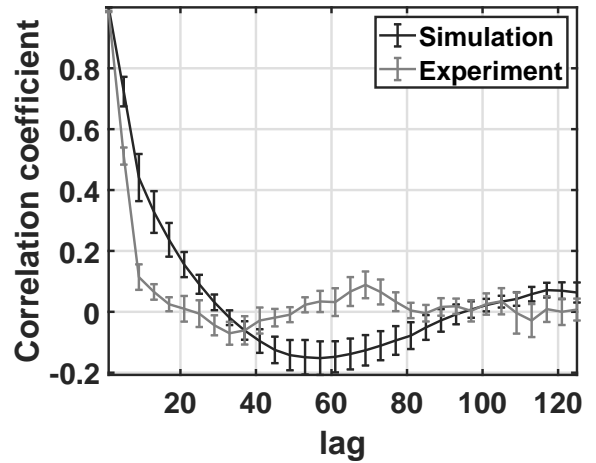}}
\end{picture}
\caption{Coherence plots comparison for simulation and experiment with errorbars for different depth positions around the focal depth of the transducer (80-110~mm), in the center of the transducer laterally. }
\label{fig:lag}
\end{figure}

Spatial coherence for the experimental and simulated data during the calibration process is measured in the same manner as in Pinton et al. (2014)~\cite{pinton2014}, namely by means of correlation of the signals received by the transducer as a function of inter-element distance, or lag as demonstrated to in Fig.~\ref{fig:lag}. The correlation coefficients were calculated for 30 different depths between 80 and 110~mm for both the experiment and simulation, which reflects in the error bars shown in Fig.~\ref{fig:lag}. The depths were selected based on the natural focal depth of the C5-2v probe for a plane wave emission, since we expect regions of higher correlation near the focal zone. Lag-one coherence (LOC), introduced by Long et al.~\cite{lag1}, can be used as an image quality metric and was found to be correlated to traditional image quality indicators such as CNR. LOC leverages the spatial coherence between nearest-neighbor array elements to provide a local measure of thermal and acoustic noise. LOC was measured at 0.91 for the experiment and 0.98 for the simulations.


\section{Results}

Average speckle brightness is calculated as the mean B-mode brightness value for the B-modes, excluding the target region. The values matched for the experiment (42~dB) and the simulation dataset (48~dB).

In Fig.~\ref{fig:bmodes} the B-modes for both experiment and simulation are labeled. The thickness of the abdomen is roughly the same at approximately 25mm, with localized reverberation apparent until roughly the depth of ~50mm.

Having established the calibration match between the real and simulated abdomen, in order to better understand the contribution of each image degrading mechanism, we can generate different datasets from the same simulation environment only containing the effects of aberration or reverberation. This is better assessed by using a full aperture focused emission, focusing directly on the scatterer region in order to achieve maximum deposition of energy.

In order to linearly isolate image degrading effects into their individual components, we generate isoimpedance and isovelocity images, To generate isovelocity images, we set the sound speed maps in the simulation to be constant at a reference value of 1570~m/s, while conserving the heterogeneous density, and attenuation maps. Non-linearity maps are kept at a zero value to allow us to linearly separate and potentially recombine the aforementioned elements. To generate isoimpedance images, we run two different simulation scenarios, one without a point scatterer and its identical counterpart including the scatterer. Subtracting the former from the latter results in an isoimpedance image, where the effects of reverberation are removed whilst preserving aberration. Additionally, we created a simulation scenario where sound speed and attenuation maps remain identical to the heterogeneous case but density is scaled accordingly for impedance to match a constant reference value equal to the homogeneous case. In ths scenario, hereon referred to as z = ct case, we notice that not only reverberation effects but also phase aberration and trailing clutter are virtually removed from the B-mode image. 

The PSF can be divided into three regions: First the isochronous volume, which is the bow-tie shaped figure centered at the focus. Signals within this volume have had time to travel from the transducer to the target and back again. The second region, referred to as preceding region, is situated above the isochronous volume which corresponds to times that precede
the arrival of the target signal, in this case multiple reverberations persisting from the abdominal wall. The third region is the trailing region of the isochronous volume which corresponds to times that follow the arrival of the target signal, such as any remainder of multiple reverberations, or trailing distortions of the ballistic pulse which constitute the trailing clutter. 

\begin{figure}[h!]
\setlength{\unitlength}{0.75\textwidth}
\begin{picture}(0.5,0.55)(0,0)
\put(0.0,0.0){\includegraphics[trim= 0 0 0 0,clip,height=0.52\unitlength]{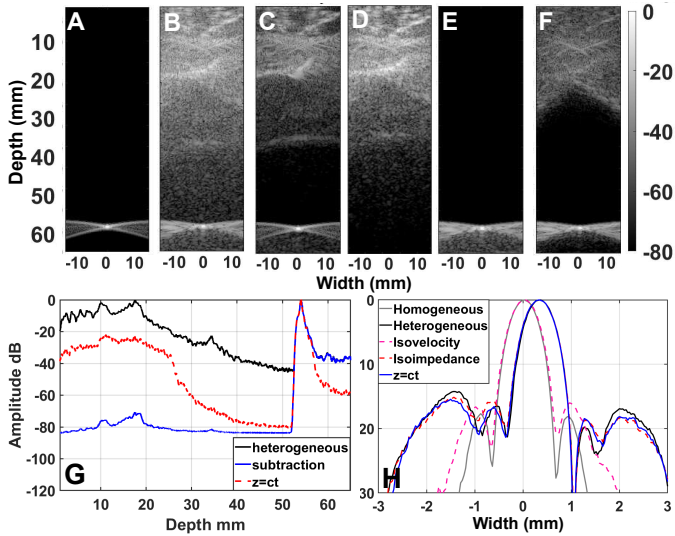}}
\end{picture}
\caption{A-E, B-mode images for a target at 55~mm from the transducer normalized by the maximum amplitude of the homogeneous case. Transducer was apodized at half the aperture. A, is the homogeneous B-mode, B is the heterogeneous B-mode through the abdomen, C is the isovelocity B-mode. D is the isompedance B-mode through clutter subtraction whereas E is the isoimpedance B-mode through setting the isoimpedance to be constant. F is the reverberation curves for three of the aforementioned cases B,D and E.}
\label{fig:psfs}
\end{figure}

The mean amplitude of each of these regions for different cases examined is summarized in Table~\ref{tab:psf}. We can note that for the preceding regions, removing aberration has the greatest impact, restoring the amplitude by 95\%, followed by isoimpedance via setting it to be globally constant, which restored the amplitude by 93\%. The addition of the abdomen in the heterogeneous case increases the trailing region amplitude at -8.6~dB, compared to -22.6~dB for the homogeneous case. The trailing region amplitude, is restored closest to its reference value when we set the impedance to be globally constant and equal to the value for the homogeneous case at -14.6~dB. Removing aberration also decreases trailing amplitude at -11.4~dB. As we mentioned, the trailing region of the PSF comprises both multiple reverberations and trailing clutter. Removing multiple reverberations via subtraction of clutter does not affect the amplitude of the trailing region, implying that multiple reverberations do not significantly contribute at this depth. Lastly, the amplitude of the isochronous regions remains relatively constant.

\begin{table}[ht]
    \centering
      \caption{Average amplitude of PSF regions for a scatterer at 5~cm in dB}
    \label{tab:psf}
    \begin{tabular}{c|c|c|c|}
         &  Preceding & Trailing & Isochronous\\
         \hline
         \hline
         Homogeneous & -23.3 & -22.6 & -22.2 \\
         \hline
         Heterogeneous & -9.2 & -8.6 & -16.7\\
         \hline
         Isoimpedance & -17.7 & -8.9 & -16.7\\
         \hline
         Isovelocity & -22.1 & -11.4 & -19\\
         \hline
         z = ct & -21.6 & -14.6 & -17\\
         \hline
    \end{tabular}
\end{table}

The results of this study can be visually summarized in Fig.~\ref{fig:psfs}. A scatterer was placed 5~cm away from the transducer in a homogeneous medium for reference, shown in Fig.~\ref{fig:psfs}A, and under an abdominal slice, cases Fig.~\ref{fig:psfs}B-D. We can qualitatively assess that aberration appears to play a crucial role in this case, whereas reverberation is secondary. In the isovelocity case, Fig.~\ref{fig:psfs}C, the point target appears to be almost fully restored to its homogeneous values.

\begin{figure}[h!]
\setlength{\unitlength}{0.75\textwidth}
\begin{picture}(0.5,0.55)(0,0)
\put(0.0,0.0){\includegraphics[trim= 0 0 0 0,clip,height=0.56\unitlength]{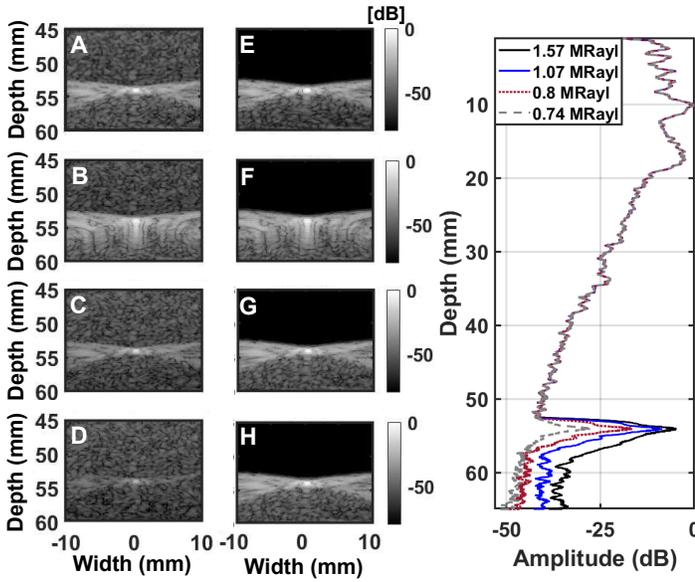}}
\end{picture}
\caption{B-mode images for impedance mismatch between target and background of: A Rayl, B, C, D. E,F,G and H are the equivalent isoimpedance via subtraction B-modes. I shows the reverberation curves for B-modes B, C and D.}
\label{fig:bright}
\end{figure}

The same study was performed for targets 5 and 8~cm away from the transducer. In all cases, we have found that removing the effects of aberration fully restores the shape of the target, as well as its original brightness value by 97.5\%. Initial aberration value for the heterogeneous case at 8~cm is 106~ns, dropping to 30~ns for the isovelocity case. For the 5~cm scatterer aberration is measured at 117~ns, which is expected since aberration does not significantly vary with depth.
As far as reverberation is concerned, removing its effects by subtracting the clutter from the RF data does not really affect the quality of the PSF at either depth. The majority of the reverberation zone spans the first 3~cm of the image, which is considered too shallow for transabdominal imaging using a transducer like the C5-2V. According to the beamplots as shown in Fig.~\ref{fig:psfs}G, with the width of the main lobe at -6~dB being 0.76~mm in the homogeneous case, 0.78~mm in the isovelocity and constant z case and 0.82~mm in the heterogeneous and isoimpedance via subtraction cases.

\begin{figure*}[h!]
\setlength{\unitlength}{0.75\textwidth}
\begin{picture}(0.9,0.4)(0,0)
\put(0.0,0.0){\includegraphics[trim= 0 0 0 0,clip,height=0.5\unitlength]{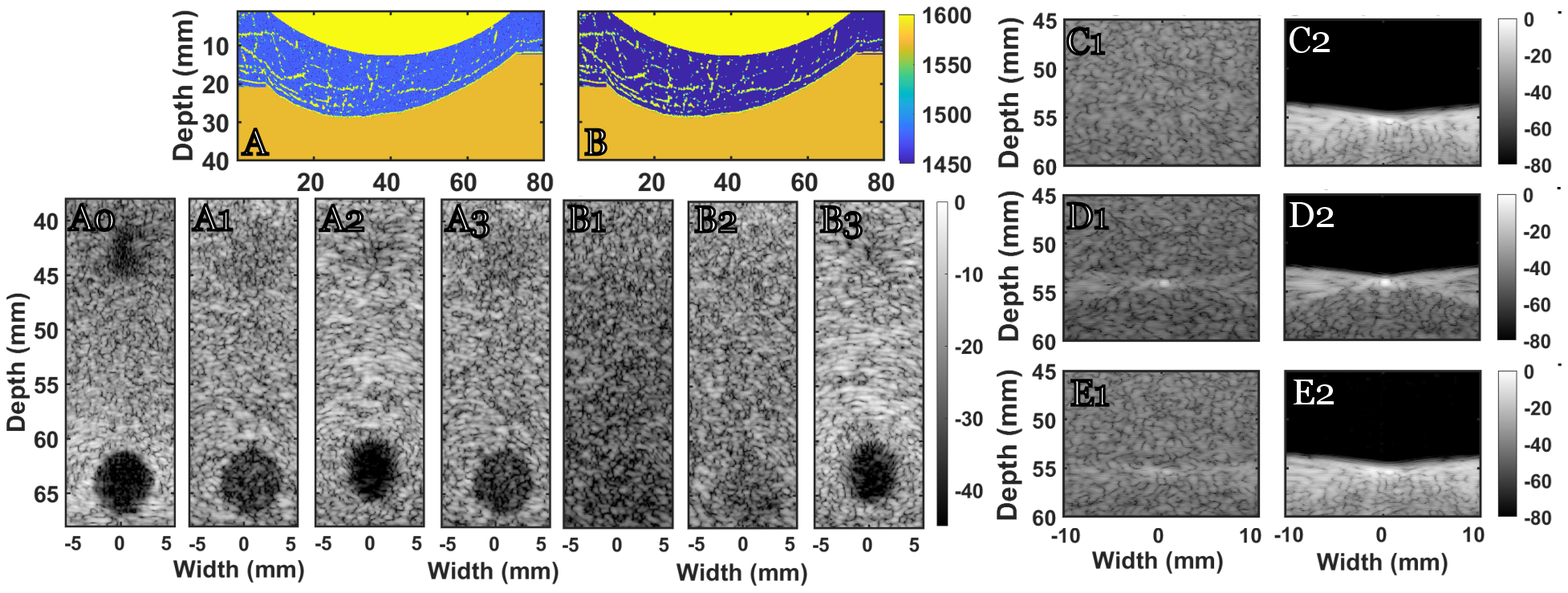}}
\end{picture}
\caption{Anechoic lesions at two different depths, 40 and 60~mm, (A1) for the homogeneous case, (A1) for a transabdominal case, (A2) isovelocity and (A3) isoimpedance corresponding to sound speed map (A). Same lesions (B1) for a transabdominal case, (B2) isoimpedance (B3) isovelocity cases corresponding to sound speed map (B), which has a 54\% larger impedance mismatch than map (A). A PSF analysis is also shown for different methods of increasing impedance mismatch, namely: (C1) Increasing speed of sound mismatch by a factor of 3 without scaling density and the equivalent isoimpedance B-mode (C2). (D1) Increasing density mismatch by a factor of 3 without sound speed and the equivalent isoimpedance B-mode (D2). (E1) Increasing sound speed mismatch by a factor of 3 while inversely scaling density and the equivalent isoimpedance B-mode (E2).}
\label{fig:lesions}
\end{figure*}

For transabdominal imaging which is traditionally aimed at greater depths, multiple reverberations do not really affect the quality of the image, since their effects are localized at very shallow regions. In a case with multiple bright scatterers however, trailing clutter could result in degradation along the direction of pulse propagation. To further investigate the effects of reverberation, we have removed using two methods, clutter subtraction and setting the global impedance to be constant and equal to the homogeneous case, both described in section II. As shown in Fig.~\ref{fig:psfs}D, subtracting the clutter from the RF data and beamforming it reduces all the reverberation effects to less than -80dB. The trailing clutter however that follows the scatterer remains exactly the same as in case Fig.~\ref{fig:psfs}B. By setting the impedance as constant (Fig.~\ref{fig:psfs}E), a portion of the effects of tissue down to 3~cm, approximately 20~dB lower than the original heterogeneous case, persist. The rest of the reverberation effects, quickly drop down to -80~dB as well, following an exponential drop after 3~cm of depth. This can be perhaps attributed to small numerical errors due to the manipulation of the density maps. As shown in Fig.~\ref{fig:psfs}E in this case, trailing clutter appears to have reduced.

In this analysis, target brightness, in this case the impedance mismatch between the scatterer and the background tissue, can also elucidate how the difference in target reflectivity can affect image degradation mechanisms. We hypothesize that phase aberration, as well as multiple reverberations from the abdomen remain relatively constant with varying reflectivity. The results of this study are shown in Fig.~\ref{fig:bright}. The same simulation setup as in Fig.~\ref{fig:psfs} was used, with only the sound speed and density values, thus impedance values, of the scatterer changing. In Fig.~\ref{fig:bright}A and E, the target impedance mismatch was set at an intermediate value of 1.07~MRayl, with A showing a zoomed in portion of the heterogeneous case and E the equivalent isoimpedance via subtraction. In B and F, target impedance was decreased to 0.096~MRayl, thus increasing the impedance mismatch between target and background to 1.57~MRayl. In C, impedance was increased to 0.87~MRayl, thus actually decreasing the mismatch to 0.8~MRayl. Lastly, mismatch was further reduced to 0.74~MRayl in D and H. As shown in Fig.~\ref{fig:bright}I, reverberation curves remain unchanged, since the target brightness is the only variable, however trailing clutter appears to scale to the impedance mismatch, with higher impedance differences generating brighter clutter. Phase aberration also remained constant, namely at RMS values of 79, 77. 78 and 79~ns respectively for E,F,G and H. Aberration values were measured from the isoimpedance versions of the PSFs in order to eliminate any effects of reverberation that degrade the profiles used to measure RMS differences.

\begin{table}[ht]
    \centering
      \caption{CNR for the two lesions for different cases.}
    \label{tab:lesion}
    \begin{tabular}{c|c|c|}
        Type of emission &  Lesion 40mm & Lesion 60mm \\
         \hline
         \hline
         Homogeneous & 3 & 4.2  \\
         \hline
         Heterogeneous & 0.6 & 1.25 \\
         \hline
         Isovelocity & 1.9 & 4.1 \\
         \hline
         Isoimpedance & 1.1 & 1.9\\
         \hline
         Higher speed abdomen & - & -\\
         \hline
         Higher speed abdomen - Isoimpedance & - & -\\
         \hline
         Higher speed abdomen - Isovelocity & 1.9 & 2.7\\
         \hline
    \end{tabular}
\end{table}

Having shown that we can successfully isolate and characterize image degradation components, we also aim to introduce additional degradation in B-mode images while providing a clinical interpretation. We therefore simulated a field of random scatterers in the speed of sound maps, at a density of 12 scatterers per resolution cell and an impedance mismatch of 6\% to the background speed. In this field we introduced two circular anechoic lesions, at depths of 40 and 60~mm respectively by removing all scatterers at a radius of 3~mm. Anechoic lesions are harder to image than a bright single target as shown in our PSF analysis due to the added effects of multiple scattering in the field. For that reason we modified our imaging sequence into a multi-focal focused wave, comprising 173 independent transmit-receive events. Each event, directs the wave to a different focus laterally, while retaining the same focal depth. The 173 individual foci, are placed at intervals of beamwidth/2, corresponding to 173~$\mu$m for the modeled transducer, and a total area of 3~cm laterally can be totally illuminated. One line per transmit is used for beamforming, resulting in the B-mode images shown in Fig.~\ref{fig:lesions}. The focal depth in all transmits is set at 65~mm, corresponding to the end of the last lesion.

After establishing a reference of the lesions in an otherwise homogeneous medium (Fig.~\ref{fig:lesions}A0), The same abdominal maps used in our PSF study, the sound speed portion of which is shown in Fig.~\ref{fig:lesions}A, to simulate a trans-abdominal emission, hereon referred to as a heterogeneous case. Illustrated in Fig.~\ref{fig:lesions}A1, the abdomen reduces both contrast and resolution in the image. Cases A2 and A3 correspond to the isovelocity and isoimpedance via clutter subtraction cases described previously. Judging by the PSF study, we hypothesize that by increasing the impedance mismatch between the abdomen and the surrounding tissue will further cause deterioration of the image, primarily by increased aberration. To examine this hypothesis the entire abdominal part was scaled linearly for both speed of sound and density maps, resulting in a cumulative impedance mismatch increase of 54\% as measured in the mean bulk of the abdomen. The heterogeneous case of this higher impedance abdomen is shown in Fig.~\ref{fig:lesions}B1, with the isoimpedance and isovelocity cases shown in B2 and B3 respectively. As we hypothesized, it is evident that removing aberration has the most effect in restoring the image back to its normal state, whereas the extra reverberation did not really make a qualitative difference in image quality.

To further quantify these results, we have measured the contrast to noise ratio (CNR) for each of the two lesions in all aforementioned cases. The results are summarized in Table ~\ref{tab:lesion}. Adding the abdominal layer reduces the CNR of the homogeneous reference by 80\% for the shallow lesion at 40~mm and 70\% for the deeper lesion at 60~mm. With the presence of the abdomen but having the effects of aberration removed, the CNR dropped by 37\% and only 2\% for the shallow and deep lesions respectively. On the other hand removing reverberation from the abdominal B-mode, results in a CNR reduction by 63 and 55\% for the two lesions respectively, showing that aberration has a far larger deleterious effect on CNR than reverberation.

When the same abdomen was subjected to a global increase in impedance mismatch as explained above, in both heterogeneous and isoimpedance cases, the lesions were practically at noise level, thus a number is not provided in Table ~\ref{tab:lesion}. Only by removing aberration we are able to discern the lesions and quantify CNR.

\section{Discussion}
In this work we have shown how the individual image degradation components can be reversibly and linearly isolated, quantified and even exaggerated in B-mode imaging through the human abdomen using the Fullwave2 simulation tool. We use the isolation into separate components to evaluate the contribution of each depending on factors such as the brightness and the depth of the target.

Aberration remains relatively constant as a function of depth as well as a function of target impedance mismatch as shown in Section II.
By linearly removing its effects from the image $in $ $silico$ we are able to assess the magnitude of the degradation by calculating the RMS difference between the target backscattered profile to a reference profile and by beamplot measurements which indicate registration errors and target distortion. In a more clinically relevant setting, we have shown that for expected values of sound speed and density mismatch, CNR values of anechoic lesions at larger depths (6~cm) are restored to 95.3\% of their original value when removing aberration. Even at a depth of 4~cm, as showcased in the case of shallow anechoic lesions, aberration seems to contribute more than half of the degradation in terms of CNR (63\% reduction in CNR).

Reverberation appears to have a hazing, blurring effect on the image that is mainly localized at shallower depths, in our case lesser than 4~cm in depth. For deeper anechoic lesions, removal of multiple reverberation via linear subtraction of clutter, resulted in an increase of 44\% of the CNR for a deep lesion, while qualitatively the shape of the lesion is not improved, as shown in Fig.~\ref{fig:lesions}. Removing reverberation was, however instrumental in ensuring the accurate measurement of aberration in ~\ref{fig:bright}, as it obscures correlation based measurements. Reverberation of the part of the image prior to the target is also constant for different values of target brightness, however, trailing clutter appears to proportionally scale as a function of impedance mismatch between the target and the background.

We also show that using this simulation method we can cover ranges of impedance mismatch far greater than those usually encountered clinically, by modifying the input maps. This flexibility along with the freedom to simulate a variety of anatomical structures is a key advantage for the generation of a large volume of datasets for purposes of algorithm training in machine learning applications. This is supported by the fact that in silico there is an inherent advantage of having a reference, or ground truth, for B-mode images, something that cannot be achieved experimentally, let alone clinically. 



\bibliographystyle{plain}
\bibliography{ref}

\end{document}